\documentclass[a4paper, 11pt]{article}

\pdfoutput=1 

\usepackage{amsmath}
\usepackage[T1]{fontenc} 
\usepackage{authblk}
\usepackage[a4paper, total={6in, 8in}]{geometry}
\usepackage{graphicx}

\begin{document} 

\title{\boldmath $SU(5)_{L} \times U(1)_{Y}$ \bf{electroweak unification}}

\author{Adrian Palcu}
\affil{"Aurel Vlaicu" University of Arad,\\2 Elena Dr\u{a}goi Street, Arad-310330, Romania}

\maketitle

\abstract{We propose here - for the first time in the literature, to our best knowledge - an electroweak unification based on the $SU(5)_{L}\times U(1)_{Y}$ gauge group. The spontaneous symmetry breaking takes place in the manner $SU(5)_{L}\times U(1)_{Y} \rightarrow U(1)_{em}$, due to a particular Higgs sector consisting of five scalar quintuplets. Each scalar quintuplet acquires its own vacuum expectation value, by means of a proper parametrization which is worked out once the overall vacuum expectation value in the model is established. The decoupling of the low energy regime (corresponding to the Standard Model) from the high scale (required by out model here) is straightforwardly achieved in order to preserve the consistency with the present experimental data. Finally, a promising phenomenological outcome is derived by simply tuning a single free parameter. Our results include, besides a viable one-parameter mass spectrum, also the prediction of precisely three generations in the fermion sector and the electric charge quantization.}

\section{Introduction}
\label{sec:1}
In a recent paper \cite{1}, the author took a step further in working out some consequences of the "general Cot\u{a}escu method" (gCm) conceived for treating $SU(3)_{c}\times SU(n)_{L}\times U(1)_{Y}$ gauge models \cite{2}, by proving that this particular approach can precisely predict the electric charge quantization observed in nature. At the same time, a promising phenomenological outcome can be achieved, once the free parameters necessary to split the scalar sector underpinning a coherent Higgs mechanism are properly set. To this end, the model must undergo a spontaneous symmetry breaking (SSB) in its electroweak sector up to the universal $U(1)_{em}$ group, while the renormalization requirement is kept valid for all the model's sectors throughout the proceedings. The main result - that allows for a plethora of new degrees of freedom beyond those coming from the Standard Model (SM) - regards the mass spectrum of the model. In our approach, the low energy regime can be satisfactorily decoupled from the overall vacuum expectation value (VEV) of such general gauge models, in order not to compromise the SM phenomenology (constructed within the well-known $SU(3)_{c}\times SU(2)_{L}\times U(1)_{Y}$ gauge group \cite{3}-\cite{5}). In this connection, a compelling argument was sketched - see section 5 in Ref.\cite{1} - for the most general case when $n$ is arbitrary. When particular values of $n$ in the gCm are taken into consideration, our approach leads to some concrete extensions of the SM. Not more than three such possible extensions \cite{1}, as the strict renormalization condition informs us. So, a vast amount of work has been published to date dealing with the so called 3-3-1 gauge models \cite{6}-\cite{32} or 3-4-1 gauge models \cite{33}-\cite{57}, but none approaching 3-5-1 model, the last one allowed by the gCm. 

But, why extending the SM? The reason to extend the SM is fueled by some recent experimental challenges - regarding, among others, neutrino oscillations or dark matter puzzle - that have definitely no explanation at the level of the SM. Instead, they seem to be promisingly addressed within the richer framework supplied by such extended gauge models. Not to add certain theoretical devices that are natural in such SM-extensions and aim at elucidating some issues (that already have come of age), such as the strong-CP problem, the electric charge quantization, the precise number of three fermion generations, the tiny (seesaw) neutrino masses accounting for the oscillation phenomenon - to name but a few. Most plausible candidates for dark matter and most suitable ingredients for various types of seesaw mechanism can be definitely found in a larger extension of the SM. That is why we address here the task of constructing the 3-5-1 model (with no exotic electric charges) based on the $SU(3)_{c}\times SU(5)_{L}\times U(1)_{Y}$ gauge group. The theoretical structure will be rigorously built assuming the gCm formalism is combined with the restrictive \emph{asymptotic freedom} requirement of the QCD \cite{58,59}. Afterwards, the consistency with the available experimental data will be highlighted. Our strategy here is to apply and effectively exploit all the prescriptions of the gCm for the particular case $n=5$.  

The paper is organized as follows. In sections \ref{sec:2} - \ref{sec:4} the three sectors (fermion, gauge, scalar) of the model are presented. In section \ref{sec:5} the electric charge quantization is obtained, while section \ref{sec:6} is devoted to some phenomenological results regarding the mass spectrum and some rough numerical estimates depending on the overall breaking scale in the model. A phenomenological outlook and related comments are stated in the final section \ref{sec:7}. Not to omit two Appendixes, where some mathematical details regarding the Lie algebra involved in the procedings and the diagonalization of the mass matrix are displayed.

\section{Fermion sector}
\label{sec:2}

We consider the simplest class of 3-5-1 gauge models, namely the one with no other electric charges than $0$, $\pm$ e, and make use of the prescriptions obtained by the author in the general case \cite{1}. Due to the asymptotic freedom imposed by the color $SU(3)$ symmetry of the QCD, there must be precisely three fermion generations in order to cancel the chiral anomalies that could occur in the electroweak sector. Therefore, the fermion sector will consist of the following left-handed (lower index L) representations (here, the generic L stands for leptons, generic Q stands for quarks, and $i=1,2,3$):

\begin{equation}
L_{iL}=\left(\begin{array}{c}
N_{i}^{\prime\prime}\\
N_{i}^{\prime}\\
N_{i}\\
\nu_{i}\\
e_{i}
\end{array}\right)_{L}, \quad Q_{1L}=\left(\begin{array}{c}
U^{\prime\prime}\\
U^{\prime}\\
U\\
u\\
d
\end{array}\right)_{L}  , \quad Q_{2L,3L}=\left(\begin{array}{c}
D_{2,3}^{\prime\prime}\\
D_{2,3}^{\prime}\\
D_{2,3}\\
d_{2,3}\\
u_{2,3}
\end{array}\right)_{L}\label{eq:2.1}
\end{equation}
along with their corresponding singlet right-handed partners. The capital letters on the first three positions in each generation are assigned to new fermions (specific to this 3-5-1 model), whilst the small letters denote the well-known SM fermions. For the moment, their electric charges are just supposed to be those observed in nature, but this will be rigorously proved only in section \ref{sec:5}.   

According to the gCm, for $n=5$ - from eqs.(35)-(37) in Ref.\cite{1} - a renormalizable model of this kind must have the fermion representations transforming under the $SU(3)_{c}\times SU(5)_{L}\times U(1)_{Y}$ gauge group in the manner:

\begin{equation}
L_{iL}\sim(\boldsymbol{1},\boldsymbol{5},-\frac{1}{5}) \quad , \quad e_{iR}(\boldsymbol{1},\boldsymbol{1},-1) \quad ,\quad \nu_{iR},N_{iR},N_{iR}^{\prime},N_{iR}^{\prime\prime}\sim(\boldsymbol{1},\boldsymbol{1},0)\label{eq:2.2}
\end{equation}

\begin{equation}
Q_{1L}\sim(\boldsymbol{3},\boldsymbol{5},\frac{7}{15}) \quad , \quad Q_{kL}\sim(\boldsymbol{3},\boldsymbol{5^{*}},-\frac{2}{15})\label{eq:2.3}
\end{equation}

\begin{equation}
u_{kR},u_{R},U_{R},U_{R}^{\prime},U_{R}^{\prime\prime}\sim(\boldsymbol{3},\boldsymbol{1},\frac{2}{3})\quad , \quad d_{R}, D_{kR},D_{kR}^{\prime},D_{kR}^{\prime\prime}\sim(\boldsymbol{3},\boldsymbol{1},-\frac{1}{3})\label{eq:2.4}
\end{equation}
with $k=2,3$.

The above assignments are sufficient in order to get a renormalizable 3-5-1 gauge model for they cancel all the chiral anomalies that could plague effective results. In addition, they will play a crucial role in the electric charge quantization.

\section{Gauge sector}
\label{sec:3}
While the QCD remains an exact $SU(3)_c$ vector symmetry, the Lie algebra of the electroweak $SU(5)_{L}\times U(1)_{Y}$ gauge group provides us with a large set of gauge interactions. In this connection, besides the SM interactions due to neutral (diagonal) bosons $A_{em}$ (massless) and $Z$ ($91.1876 \pm 0.0021$ GeV \cite{60}), three more neutral weak interactions arise as coupled to heavier (also diagonal, Hermitian) bosons $Z^{\prime}$,$Z^{\prime\prime}$ and $Z^{\prime\prime\prime}$, respectively. At the same time, there are 18 more off-diagonal (non-Hermitian) entries in the adjoint representation, other than $W^{\pm}$ ($80.379 \pm 0.012$ GeV \cite{60}). Some of these 18 are charged, some are neutral - as it will be made clear further on. The electroweak boson adjoint representation stands as

\begin{equation}
A_{\mu}=\left(\begin{array}{ccccccccc}
D_{\mu}^{1} &  & Y_{\mu}^{\prime\prime0} &  & Y_{\mu}^{\prime0} &  & Y_{\mu}^{0} &  & Y_{\mu}^{\prime\prime+}\\
\\
Y_{\mu}^{\prime\prime0*} &  & D_{\mu}^{2} &  & X_{\mu}^{\prime\prime0} &  & X_{\mu}^{\prime0} &  & Y_{\mu}^{\prime+}\\
\\
Y_{\mu}^{\prime0*} &  & X_{\mu}^{\prime\prime0*} &  & D_{\mu}^{3} &  & X_{\mu}^{0} &  & Y_{\mu}^{+}\\
\\
Y_{\mu}^{0*} &  & X_{\mu}^{\prime0*} &  & X_{\mu}^{0*} &  & D_{\mu}^{4} &  & W_{\mu}^{+}\\
\\
Y_{\mu}^{\prime\prime-} &  & Y_{\mu}^{\prime-} &  & Y_{\mu}^{-} &  & W_{\mu}^{-} &  & D_{\mu}^{5}
\end{array}\right)\label{eq:3.1}
\end{equation}
where the diagonal entries are in order:
 
\begin{equation}
\begin{array}{ll}
D_{\mu}^{1} =&  \frac{1}{2}A_{\mu}^{3}+\frac{1}{2\sqrt{3}}A_{\mu}^{8}+\frac{1}{2\sqrt{6}}A_{\mu}^{15}+\frac{1}{2\sqrt{10}}A_{\mu}^{24}+YB^{0}_{\mu} \\
\\
D_{\mu}^{2} =  &  -\frac{1}{2}A_{\mu}^{3}+\frac{1}{2\sqrt{3}}A_{\mu}^{8}+\frac{1}{2\sqrt{6}}A_{\mu}^{15}+\frac{1}{2\sqrt{10}}A_{\mu}^{24}+YB^{0}_{\mu} \\
\\
D_{\mu}^{3} = &  -\frac{1}{\sqrt{3}}A_{\mu}^{8}+\frac{1}{2\sqrt{6}}A_{\mu}^{15}+\frac{1}{2\sqrt{10}}A_{\mu}^{24}+YB^{0}_{\mu} \\
\\
D_{\mu}^{4} = &   -\frac{3}{2\sqrt{6}}A_{\mu}^{15}+\frac{1}{2\sqrt{10}}A_{\mu}^{24}+YB^{0}_{\mu} \\
\\
D_{\mu}^{5} = & -\frac{2}{\sqrt{10}}A_{\mu}^{24}+YB^{0}_{\mu}
\end{array}\label{eq:3.2}
\end{equation} 

The off-diagonal entires are generically denoted as $B_{\mu}^{\alpha\beta}=\frac{1}{\sqrt{2}}(A_{\mu}^{\alpha}\pm iA_{\mu}^{\beta})$ with $\alpha,\beta=1,2,3,4,5$, $\alpha\neq\beta$. Some of the off-diagonal entries could correspond to charged bosons (if either $\alpha=5$ or $\beta=5$), but also to neutral bosons as well (if simultaneously $\alpha\neq5$ and $\beta\neq5$) - as one will be able to infer, once the charge quantization will be performed in the fermion sector. However, even at this stage, one can state (based only on algebraical arguments) that $B_{\mu}^{\alpha\beta}=B_{\mu}^{*\beta\alpha}$.

\section{Scalar sector}
\label{sec:4}
In order for the model to achieve its SSB, there are needed no less than five quintuplets, each quintuplet acquiring its own VEV, according to a proper prametrization \cite{1,2} that will ensure the VEVs split. Before the SSB, the scalar quintuplets transform like

\begin{equation}
\phi^{(i)}=\left(\begin{array}{l}
\phi_{1}^{(i)}\\
\phi_{2}^{(i)}\\
\phi_{3}^{(i)}\\
\phi_{4}^{(i)}\\
\phi_{5}^{(i)}
\end{array}\right)\sim(\boldsymbol{1},\boldsymbol{5},Y_{i})\qquad i=1,2,\ldots,5\label{eq:4.1}
\end{equation}
with the generic $Y_{i}$ the would-be hypercharge associated to $U(1)_{Y}$. As the gCm restricts \cite{2}, these would-be hypercharges must obey Tr($Y$)=0 for the scalar multiplets, namely $Y_{1}+Y_{2}+Y_{3}+Y_{4}+Y_{5}=0$.

After the SSB, each scalar quintuplet will exhibit its own VEV  

\begin{equation}
\left\langle \phi^{(i)}\right\rangle =\left(\begin{array}{c}
\vdots\\
\eta_{i}\left\langle \varphi\right\rangle \\
\vdots
\end{array}\right)\qquad i=1,2,\ldots,5\label{eq:4.2}
\end{equation}
but at least one or two of them must range in the hundreds of GeV to recover the SM phenomenology and prevent it from being altered by the high energy scale in the model. The remaining VEVs, of course, are ranging higher and they will depend on the free parameters $\eta_{i}$. The general solution for a good renormalizable model predicts - eqs. (33)-(34) in Ref.\cite{1} - precisely for the fifth quintuplet $Y_{5}=\frac{4}{5}$ and for the first four ones $Y_{k}=-\frac{1}{5}$, $k=1,2,\ldots,4$. 

Now, what are the parameters $\eta_{i}$ and why do we need them? They are key ingredients in the gCm - section 4 in Ref.\cite{2} - in order to split the VEVs in the manner $\eta_{i}\left\langle \varphi\right\rangle $ for each $i=1,2,\ldots,n$, once the overall VEV in the model $\left\langle \varphi\right\rangle $ is established via the minimum condition imposed to the renormalizable scalar potential allowed by the gauge symmetry (without any other supplementary restrictions). Obviously $\eta_{i}\in(0,1)$. The parameter matrix involved in the scalar sector looks like

\begin{equation}
\eta=\left(\sqrt{1-\eta_{0}^{2}}\right)\textrm{Diag}(\eta_{1},\eta_{2},\eta_{3},\eta_{4},\eta_{5})\label{eq:4.3}
\end{equation}
with $\eta_{0}$ the corresponding coefficient of $\varphi\sim$ ($\bf{1}$, $\bf{1}$, 0) in the kinetic term of the potential (eq.(30) in Ref.\cite{2}). Actually, this will be the unique scalar field predicted by the gCM that, in turn, will split into the five surviving Higgses $h_{i}=\eta_{i}\varphi$ , after SSB takes place.   

The compulsory trace condition $Tr(\eta^{2})=1-\eta_{0}^{2}$  is simply achieved as a sum of the parameters

\begin{equation}
\eta_{1}^{2}+\eta_{2}^{2}+\eta_{3}^{2}+\eta_{4}^{2}+\eta_{5}^{2}=1\label{eq:4.4}
\end{equation}

Hence, we have

\begin{equation}
\left\langle \phi_{1}\right\rangle ^{2}+\left\langle \phi_{2}\right\rangle ^{2}+\left\langle \phi_{3}\right\rangle ^{2}+\left\langle \phi_{4}\right\rangle ^{2}+\left\langle \phi_{5}\right\rangle ^{2}=\left\langle \varphi\right\rangle ^{2}\label{eq:4.5}
\end{equation}

In our particular 3-5-1 case, we propose, for the sake of simplicity, the following plausible parametrization  

\begin{equation}
\left\langle \phi_{1}\right\rangle \simeq\left\langle \phi_{2}\right\rangle \simeq\left\langle \phi_{3}\right\rangle =\sqrt{\frac{1-a}{3}}\left\langle \varphi\right\rangle \label{eq:4.6}
\end{equation}

\begin{equation}
\left\langle \phi_{4}\right\rangle =\sqrt{\frac{a-b}{2}}\left\langle \varphi\right\rangle \quad , \quad \left\langle \phi_{5}\right\rangle =\sqrt{\frac{a+b}{2}}\left\langle \varphi\right\rangle \label{eq:4.7}
\end{equation}
that, obviously, satisfies eq.\eqref{eq:4.4}. 

At the end of the proceedings, the desired phenomenology is obtained if one tunes the parameters to be small enough, i.e. $a,b\rightarrow0$. Under these circumstances, the quite degenerate VEVs $\left\langle \phi_{1}\right\rangle \simeq\left\langle \phi_{2}\right\rangle \simeq\left\langle \phi_{3}\right\rangle$  range similarly to the overall VEV $\left\langle \varphi\right\rangle$, while $\left\langle \phi_{4}\right\rangle$ and $\left\langle \phi_{5}\right\rangle$ are responsible for the low energy regime (precisely corresponding to the SM phenomenology). The assumption of the strict equality for the first three VEVs is taken just for simplicity. At any time one can perform some small deviations from it in order to split the three VEVs, but this will not shift them from the high energy regime. It will only add some complications in the high energy regime, due to the mixing among heavier degrees of freedom that will arise.   

\section{Electric charge quantization}
\label{sec:5}
The expression of the electric charge operator established in the general method (eq.(58) in Ref.\cite{2}) was worked out by the author in the case of general models with no exotic electric charges (eq.(9) in Ref.\cite{1}). Hence, for our particular 3-5-1 model of interest here, the electric charge operator, when considering $n=5$, yields:

\begin{equation}
Q_{em}^{\rho}=g\sin\theta_{W}\left[T_{24}^{\rho}\frac{\sin\theta}{\sin\theta_{W}}+\frac{g^{\prime}}{g}Y^{\rho}\frac{\cos\theta}{\sin\theta_{W}}\right]\label{eq:5.1}
\end{equation}
where $\varrho$ labels each representation of the gauge group. It
will take the particular form
\begin{equation}
\frac{Q_{em}^{\rho}}{e}=\frac{1}{2\sqrt{10}}\left(\begin{array}{ccccc}
1\\
 & 1\\
 &  & 1\\
 &  &  & 1\\
 &  &  &  & -4
\end{array}\right)\frac{\sin\theta}{\sin\theta_{W}}+\frac{g^{\prime}}{g}Y^{\rho}\frac{\cos\theta}{\sin\theta_{W}}\label{eq:5.2}
\end{equation}

Once the relation between the angles $\theta$ (of the gCm) and $\theta_{W}$ (of the SM) is established - via eq.(10) in Ref.\cite{1} - it gets its 3-5-1 specific form, namely

\begin{equation}
\sin\theta=2\sqrt{\frac{2}{5}}\sin\theta_{W}\label{eq:5.3}
\end{equation}
and supplies the following formula for the couplings match

\begin{equation}
g=g^{\prime}\frac{\sqrt{5}\sin\theta_{W}}{\sqrt{5-8\sin^{2}\theta_{W}}}\label{eq:5.4}
\end{equation}

Consequently, the final expression of the electric charge operator can be easily inferred by combining the previous three equations. It becomes:

\begin{equation}
\frac{Q_{em}^{\rho}}{e}=2\sqrt{\frac{2}{5}}T_{24}^{\rho}+Y^{\rho}\label{eq:5.5}
\end{equation}

The elementary electric charge now can be obtained in terms of coupling constants of the groups involved in the electroweak unification 

\begin{equation}
\frac{\sqrt{5}g^{\prime}g}{\sqrt{5g^{2}+8g^{\prime2}}}=e\label{eq:5.6}
\end{equation}

Under these conditions, the electric charge operator eq.\eqref{eq:5.5} leads straightforwardly to the observed quatization for each representation in eqs. \eqref{eq:2.2}-\eqref{eq:2.4}, once their would-be hypercharges are plugged in. For the sake of completeness, we display below the eigenvalues of the electric charge operator for all fermion representationsthat are in order: 

\begin{equation}
Q_{em}^{(\boldsymbol{5},-\frac{1}{5})}=\left(\begin{array}{ccccc}
0\\
 & 0\\
 &  & 0\\
 &  &  & 0\\
 &  &  &  & -1
\end{array}\right)\label{eq:5.7}
\end{equation}
for the lepton sector, and

\begin{equation}
Q_{em}^{(\boldsymbol{5},\frac{7}{15})}=\left(\begin{array}{ccccc}
2/3\\
 & 2/3\\
 &  & 2/3\\
 &  &  & 2/3\\
 &  &  &  & -1/3
\end{array}\right)\label{eq:5.8}
\end{equation}

\begin{equation}
Q_{em}^{(\boldsymbol{5^{*}},-\frac{2}{15})}=\left(\begin{array}{ccccc}
-1/3\\
 & -1/3\\
 &  & -1/3\\
 &  &  & -1/3\\
 &  &  &  & 2/3
\end{array}\right)\label{eq:5.9}
\end{equation}
for the quark sector, respectively.

The trace condition for the electric charge operator $\sum_{\varrho}\textrm{Tr}(Q^{\varrho})=0$ in any good phenomenological model is fulfilled by this particular outcome, as one can easily check out. For a viable phenomenology it is obvious that the first three entries in each fermion quintuplet must be much heavier than their partners on the 4th and 5th position, since the latter ones are the well-known SM fermions. However, the fermion mass spectrum will be utterly obtained once a proper Yukawa sector is employed, but this lies beyond the scope of our paper.

\section{Boson mass spectrum}
\label{sec:6}
The next step towards a good phenomenological model is to obtain a viable mass spectrum for the boson sector. The latter - with the above resulting quantized charges - evidently, will consist only of neutral and singly-charged physical bosons. 

According to the general prescription in Ref.\cite{2}, the off-diagonal entries in eq.\eqref{eq:3.1} aquire the following masses:

\begin{equation}
M_{j}^{i}=\frac{g}{2}\sqrt{1-\eta_{0}^{2}}\left\langle \varphi\right\rangle \sqrt{\eta_{i}^{2}+\eta_{j}^{2}}\label{eq:6.1}
\end{equation}

The lighter of them (corresponding to the 4-5 position)

\begin{equation}
M^{2}(W_{\mu}^{\pm})=\frac{1}{4}g^{2}(1-\eta_{0}^{2})\left\langle \varphi\right\rangle ^{2}a\label{eq:6.2}
\end{equation}
corresponds to the well-known $W^{\pm}$ from the SM. In order to simplify expressions, one can adopt the notation $m^{2}=\frac{1}{4}g^{2}(1-\eta_{0}^{2})\left\langle \varphi\right\rangle ^{2}$.
Furthermore, the other bosons masses are:

\begin{equation}
M^{2}(Y_{\mu}^{\pm},Y_{\mu}^{\prime\pm},Y_{\mu}^{\prime\prime\pm})= m^{2}\left(\frac{1-a}{3}+\frac{a+b}{2}\right)=\frac{M^{2}(W_{\mu}^{\pm})}{a}\left(\frac{1}{3}+\frac{a+3b}{6}\right)\label{eq:6.3}
\end{equation}

\begin{equation}
M^{2}(Y{}_{\mu}^{0},Y_{\mu}^{\prime0},Y_{\mu}^{\prime\prime0}) =m^{2}\left(\frac{1-a}{3}+\frac{a-b}{2}\right)=\frac{M^{2}(W_{\mu}^{\pm})}{a}\left(\frac{1}{3}+\frac{a-3b}{6}\right)\label{eq:6.4}
\end{equation}

\begin{equation}
M^{2}(X_{\mu}^{0},X_{\mu}^{\prime0},X_{\mu}^{\prime\prime0})=m^{2}\left(\frac{2-a}{3}\right)=\frac{M^{2}(W_{\mu}^{\pm})}{a}\left(\frac{2}{3}\right)(1-a)
\label{eq:6.5}
\end{equation}

For the diagonal (Hermitian) bosons, i.e. the neutral weak mediators $Z,Z^{\prime},Z^{\prime\prime},Z^{\prime\prime\prime}$, the gCm - eqs.(53)-(54) in Ref.\cite{2} - provides us with the mass matrix:

\begin{equation}
M_{\hat{i}\hat{j}}^{2}=\left\langle \varphi\right\rangle ^{2}Tr(B_{\hat{i}}B_{\hat{j}})\label{eq:6.6}
\end{equation}
where 
\begin{equation}
B_{\hat{i}}=g\left[D_{\hat{i}}+\nu_{\hat{i}}(D\nu)\frac{1-\cos\theta}{\cos\theta}\right]\eta\label{eq:6.7}
\end{equation}
with $\hat{i},\hat{j}$ labeling the diagonal generators of the Cartan
subalgebra, \emph{i.e.} $\hat{i}=1,...,4$, in our particular case. Versors $\nu_{\hat{i}}=\nu^{\hat{i}}$
with $\nu_{\hat{i}}\nu^{\hat{i}}=1$ are associated to the orthogonal directions on the electromagnetic direction in the parameter space. Around $\nu$ is performed the $SO(n)$ rotation of angle $\theta$ that separates the massless photon (see eqs. (48)-(49) in \cite{2}) from the massive $Z$s. The choice of these versors actually discriminates among different models. As we work in the no-exotic-electric-charges model, the versor choice is obviously $\nu_{n^{2}-1}=1$, while all the others vanish $\nu_{\hat{i}}=0$ for $\hat{i}=1,2,3$. Actually, we have already exploited this choice when writing down the electric charge operator for no-exotic models. 

Under these circumstances, the diagonal entries of the neutral bosons matrix read

\begin{equation}
M_{11}^{2}=M_{22}^{2}=m^{2}\left(\frac{2}{3}\right)(1-a)\label{eq:6.8}
\end{equation}

\begin{equation}
M_{33}^{2}=\frac{m^{2}}{6}\left(1+\frac{7}{2}a-\frac{9}{2}b\right)\label{eq:6.9}
\end{equation}

\begin{equation}
M_{44}^{2}=\frac{m^{2}}{10\cos^{2}\theta}\left(1+\frac{15}{2}a+\frac{15}{2}b\right)\label{eq:6.10}
\end{equation}

For the first two neutral bosons one gets - as expected - no mixing. The masses of the heavier $Z^{\prime\prime}$ and $Z^{\prime\prime\prime}$ are thus degenerate.

\begin{equation}
M_{12}^{2}\simeq M_{21}^{2}\simeq M_{13}^{2}\simeq M_{31}^{2}\simeq M_{23}^{2}\simeq M_{32}^{2}=0\label{eq:6.11}
\end{equation}

There is however one exception that occurs between the last two bosons, $Z$ and $Z^{\prime}$, that notwithstanding mix with one another. 

\begin{equation}
M_{34}^{2}=M_{43}^{2}\simeq\frac{m^{2}}{2\sqrt{15}\cos\theta}\left(1-\frac{5}{2}a+\frac{3}{2}b\right)\label{eq:6.12}
\end{equation}

Therefore, one must perform the diagonalization proceedure on the matrix $M^{2}(Z,Z^{\prime})$ that has the following expression:

\begin{equation}
M^{2}(Z,Z^{\prime})=\frac{m^{2}}{2}\left(\begin{array}{cc}
\frac{1}{3}\left(1+\frac{7}{2}a-\frac{9}{2}b\right) & \frac{1}{\sqrt{15}\cos\theta}\left(1-\frac{5}{2}a+\frac{3}{2}b\right)\\
\\
\frac{1}{\sqrt{15}\cos\theta}\left(1-\frac{5}{2}a+\frac{3}{2}b\right) & \frac{1}{5\cos^{2}\theta}\left(1+\frac{15}{2}a+\frac{15}{2}b\right)
\end{array}\right)\label{eq:6.13}
\end{equation}

By imposing that one of the eigenvalues of this symmetric matrix must be

\begin{equation}
M^{2}(Z)=\frac{M^{2}(W_{\mu}^{\pm})}{\cos^{2}\theta_{W}}=\frac{m^{2}a}{\cos^{2}\theta_{W}}\label{eq:6.14}
\end{equation}
in accordance with the established SM condition, and exploiting that in our 3-5-1 model $\cos\theta=\sqrt{1-\frac{8}{5}\sin^{2}\theta_{W}}$ holds, one gets through the diagonalization procedure (see Appendix) the following relation between parameters

\begin{equation}
b=-a\tan^{2}\theta_{W}\label{eq:6.15}
\end{equation}

Consequently, the mass of the $Z^{\prime}$ will be

\begin{equation}
M^{2}(Z^{\prime})= m^{2}\left[\frac{4}{3}\left(\frac{1-\sin^{2}\theta_{W}}{5-8\sin^{2}\theta_{W}}\right)-\frac{a}{3}\frac{(5-10\sin^{2}\theta_{W}-4\sin^{4}\theta_{W})}{(5-8\sin^{2}\theta_{W})(1-\sin^{2}\theta_{W})}\right]\label{eq:6.16}
\end{equation}

At this point, the full boson mass spectrum depending on the unique free parameter $a$ is fully determined. It remains only to be tuned in order to keep consistency with the experimental data. With the above formulas - eqs \eqref{eq:6.3}-\eqref{eq:6.5}, \eqref{eq:6.8} and \eqref{eq:6.16} - the analytic expressions will take the shape:

\begin{equation}
M(Y_{\mu}^{\pm},Y_{\mu}^{\prime\pm},Y_{\mu}^{\prime\prime\pm})=\frac{M(W_{\mu}^{\pm})}{\sqrt{a}}\sqrt{\frac{1}{3}+a\left(\frac{1-3\tan^2\theta_W}{6}\right)}\label{eq:6.17}
\end{equation}

\begin{equation}
M(Y{}_{\mu}^{0},Y_{\mu}^{\prime0},Y_{\mu}^{\prime\prime0})=\frac{M(W_{\mu}^{\pm})}{\sqrt{a}}\sqrt{\frac{1}{3}+a\left(\frac{1+3\tan^2\theta_W}{6}\right)}\label{eq:6.18}
\end{equation}

\begin{equation}
M(X_{\mu}^{0},X_{\mu}^{\prime0},X_{\mu}^{\prime\prime0},Z^{\prime\prime},Z^{\prime\prime\prime})=\frac{M(W_{\mu}^{\pm})}{\sqrt{a}}\sqrt{\left(\frac{2}{3}\right)(1-a)}\label{eq:6.19}
\end{equation}

\begin{equation}
M(Z^{\prime})=\frac{M(W_{\mu}^{\pm})}{\sqrt{a}}\sqrt{(0.322-0.343a)}\label{eq:6.20}
\end{equation}

\begin{table}[h!]
\caption{Boson masses at various breaking scales (in TeV)}
\label{tab:1}       
\begin{tabular}{|l|cccccc|}
\hline\noalign{\smallskip}
{\bf mass\textbackslash} $\left\langle
\varphi\right\rangle$ & 1TeV & 2TeV & 7TeV  & 10TeV  & 13TeV & 14TeV\\ 
\noalign{\smallskip}
\hline\noalign{\smallskip}
$W^\pm$ & 80.38GeV & 80.38GeV &  80.38GeV & 80.38GeV & 80.38GeV & 80.38GeV \\
$Z^0$ & 91.18GeV & 91.18GeV &  91.18GeV & 91.18GeV & 91.18GeV & 91.18GeV \\
$Z'^0$ & 180.2GeV & 368.3GeV &  1.297TeV & 1.854TeV & 2.452TeV & 2.653TeV \\
$Y^\pm, Y'^\pm, Y''^\pm$ & 189.8GeV  & 377.7GeV &  1.320TeV & 1.886TeV & 2.452TeV & 2.653TeV \\
$Y^0, Y'^0, Y''^0$ & 194.6GeV  & 380.2GeV &  1.320TeV & 1.886TeV & 2.452TeV & 2.653TeV  \\
$X^0, X'^0,X''^0$ & 259.9GeV  & 530.3GeV &  1.867TeV & 2.668TeV & 3.469TeV & 3.740TeV \\
$Z''^0, Z'''^0$ & 259.9GeV  & 530.3GeV &  1.867TeV & 2.668TeV & 3.469TeV & 3.740TeV \\
\noalign{\smallskip}\hline
\end{tabular}
\end{table}
Some numerical estimates for the boson masses - at various breaking scales defined by the parameter $a$ - are inferred and displayed in Table 1.

To each $a\in(0,1)$ there is a corresponding breaking scale in the model, according to $\left\langle \varphi\right\rangle_{SM}=\sqrt{a}\left\langle \varphi\right\rangle$. This is straightforwardly obtained if one simply equates the mass of the $W^\pm$ boson, both in the SM and in our 3-5-1 model, with $\left\langle\varphi\right\rangle_{SM}=246$GeV.

At the same time, one can plot the boson mass spectrum (virtually, for all values of the parameter $a$). We confine however, in Figure 1, to displaying it only for a particular domain from $(0,1)$ where the high scale is clearly decoupled from the low energy regime. This region favors, as expected, very small $a$.

\begin{figure}[h!]
\includegraphics[width=\linewidth]{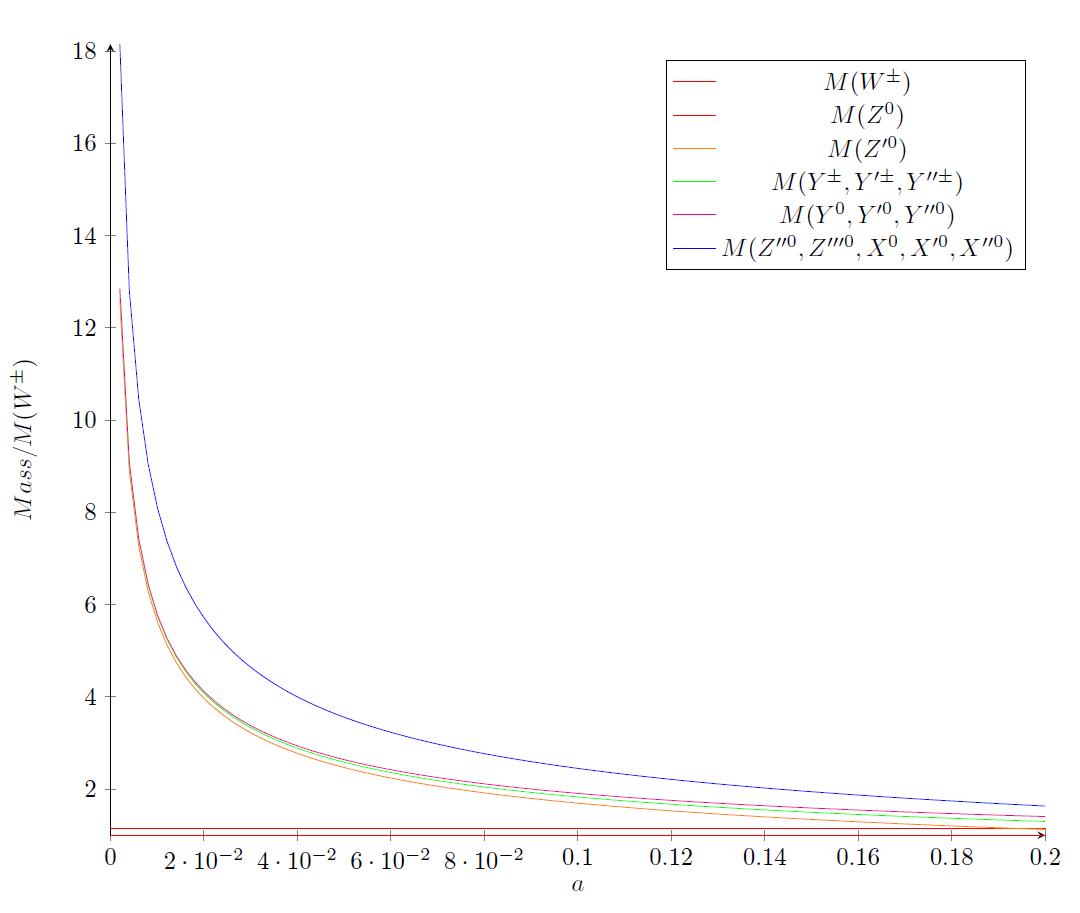}
\caption{Boson masses depending on the free parameter a}
\label{fig:1}       
\end{figure}

\section{Summary and outlook}
\label{sec:7}
In this article we set up a plausible electroweak unification by taking into consideration for that purpose the particular $SU(5)_{L}\times U(1)_{Y}$ gauge group. Within this framework, we applied rigorously the prescriptions and results of the the theoretical device conceived by Cot\u{a}escu for $SU(n)_{L}\times U(1)_{Y}$ gauge models with SSB \cite{1,2}. Once the model's symmetry was broken up to the universal electromagnetic $U(1)_{em}$, the method makes room for a viable boson mass spectrum by tuning a single free parameter $a$. Among the appealing features inherited from the general method, we have here the exact predictions for the number of fermion generations and the electric charge quantization. 

Henceforth, one can work out some more phenomenological consequences that our 3-5-1 model is suitable for. We briefly review some of them here:

\begin{itemize}
\item {\bf Weak currents}. The main next step in exploiting our theoretical approach from a phenomenological standpoint, is to determine the exact couplings of all the weak currents - properly speaking, to get the quantization of the "neutral charges" $Q^{\rho}(Z_{\hat{i}})$. That is, one has to compute the following diagonal operators (eq.(59) in Ref.\cite{2})

\begin{equation}
Q^{\rho}(Z_{\hat{i}})=g\left[D_{\hat{k}}^{\rho}-\nu_{\hat{k}}\left(D^{\varrho}\nu\right)(1-\cos\theta)-\nu_{\hat{k}}\frac{g^{\prime}}{g}Y^{\rho}\sin\theta\right]\omega_{\cdot\hat{i}}^{\hat{k}\cdot}\label{eq:7.1}
\end{equation}
for the particular 3-5-1 model at hand. Moreover, the eigenvalues of these operators must be obtained. Based on the resulting weak charges, a huge amount of phenomenology can be computed and put eventually under test at LHC and future facilities. 

To this end, one must first determine the $SO(4)$ rotation that properly diagonalizes the mass matrix \eqref{eq:6.13}. In our model, assuming the degenerate heavier $Z^{\prime\prime}$ and $Z^{\prime\prime\prime}$ bosons, it can be put as

\begin{equation}
\omega=\left(\begin{array}{cccc} 1 & 0 & 0 & 0 \\
0 & 1 & 0 & 0 \\
0 & 0 & \cos\phi & \sin\phi \\
0 & 0 & -\sin\phi & \cos\phi \end{array} \right)\label{eq:7.2}
\end{equation}

Under these circumstances, the neutral currents in the 3-5-1 model, read:

\begin{equation}
Q^{\rho}(Z)=g\left[T_{15}^{\rho}\cos\phi+\left(T_{24}^{\varrho}\cos\theta-\frac{g^{\prime}}{g}Y^{\rho}\sin\theta\right)\sin\phi\right]\label{eq:7.3}
\end{equation}

\begin{equation}
Q^{\rho}(Z^{\prime})=g\left[-T_{15}^{\rho}\sin\phi+\left(T_{24}^{\varrho}\cos\theta-\frac{g^{\prime}}{g}Y^{\rho}\sin\theta\right)\cos\phi\right]\label{eq:7.4}
\end{equation}

\begin{equation}
Q^{\rho}(Z^{\prime\prime})=gT_{8}=g\frac{1}{2\sqrt{3}}\textrm{Diag}(1,1,-2,0,0)\label{eq:7.5}
\end{equation}

\begin{equation}
Q^{\rho}(Z^{\prime\prime\prime})=gT_{3}=g\frac{1}{2}\textrm{Diag}(1,-1,0,0,0)\label{eq:7.6}
\end{equation}
 
Obviously, when inspecting eqs. \eqref{eq:7.3}-\eqref{eq:7.6}, one observes that SM fermions couple only with $Z$ and $Z^{\prime}$ being, thus, sterile with respect to the two heavier $Z^{\prime\prime}$ and $Z^{\prime\prime\prime}$. This is a good phenomenological outcome since, to date, the occurrence of $Z^{\prime\prime}$ and $Z^{\prime\prime\prime}$ has not been  reported at the available energies the present facilities can supply. A detailed determination of all the couplings remains be performed elsewhere \cite{61}, since it lies beyond the scope of our present paper. Yet, we can state that all the couplings of the ordinary particles (denoted by small letters leptons and quarks in \eqref{eq:2.1}-\eqref{eq:2.4}) recover - by computing  \eqref{eq:7.3} - precisely their SM-couplings. So, our SM-extension doesn't alter at all the SM phenomenology.     

\item {\bf Neutrino mass}. The phenomenon of neutrino oscillation (observed for the first time by Kamiokande \cite{62}  and SNO \cite{63} collaborations) requires tiny neutrino masses for the three $\nu_{iL}$. As long as a fermion mass term presumes the existence of both left-handed and right-handed chiral components, the task of directly generating such non-zero Dirac masses for neutrinos is unbearable within the SM, since no $\nu_{iR}$ can be found therein. Even if they would have been added by hand, a non natural fine tuning of their Yukawa couplings must definitely be imposed. Among many proposals for a way out of this conundrum, some seesaw mechanisms \cite{64}-\cite{70} were conceived to generate tiny Majorana masses for $\nu_{iL}$, but still resorting to an extra-SM hypothesis: the existence of some sterile heavy right-handed Majorana neutrinos ($N_{iR}$) violating the $B-L$ symmetry at a scale much higher than the SM scale. Meanwhile, many versions of the seesaw mechanism have been unfolded. Our model is flexible enough in order to supply many ingredients to such a goal and accommodate a lot of seesaw mechanisms. First of all, since all right-handed neutrinos in our model are singlets belongng to the representation  $({\bf 1}, {\bf 1}, 0)$ any of them can be removed from the spectrum without altering the cancellation of the chiral anomaly necessary to preserve the renormalizability of the model. This allows one to conceive various seesaw scenarios. Some of the massive neutral particles included in lepton quintuplets may acquire their Majorana mass through an effective dimension-5 operator approach if one takes into consideration such terms
\begin{equation}
-\frac{Y^{ij}}{\Lambda}\overline{L}L(\phi_{i}\times\phi_{j})
\end{equation} 
with $i,j=1,2,3$. However, even the radiative mass generation scenarios are not excluded in the 3-5-1 framework, but all these deserves a distinct study in its own right.     

\item {\bf Dark Matter}. The cumbersome issue of the presence of the Dark Matter in the universe can be addressed within this kind of  models, since there is a set of several particles (fermions, scalars, even neutral gauge bosons) that are sterile when it comes to the ordinary matter. Therefore they could play the role of WIMP particles and their relic abundance in the universe can, in principle, be estimated once the overall breaking scale is established. In particular 3-3-1 models there are several attempts to propose DM candidates \cite{71}-\cite{79}. If one proceeds with our simple 3-5-1 model in which the three heavier breaking scales are degenerate, for instance all the three degenerate heavy neutral fermions could be suitable cold DM candidates. A detailed discussion in this respect deserves to be separately made too.     

\item {\bf Higgs sector}. It is evident that the scalar sector of such 3-5-1 model are compounded by a plethora of scalar degrees of freedom, five of the neutral ones developing their proper VEVs and becoming Higgs particles to be hunted at LHC. A detailed analysis of the scalar spectrum could supply interesting results with regard to their masses and couplings. In the case of 3-3-1 models with no exotic electric charges, the author succeded in proving that only two physical Higgses survive the SSB \cite{18,32}, while the third one comes out massless and sterile (with no couplings at all). Some of them could also be put forward as plausible DM candidates.
\end{itemize}

\appendix
\section{{Lie algebra $su(5)_{L}\times u(1)_{Y}$}}

We briefly present the unitary irreducible representations (irreps) and  algebraical structure of the semi-simple $SU(5)\times U(1)$ Lie group with a focus on its Hermitian generators, of interest in constructing the adjoint representation defining the gauge bosons of the model.

For $U(1)_{Y}$ group things are simple, as its unitary irreps are constructed just as phase factors of its associated Lie algebra, once the single Hermitian generator is identified with $I$ and $g^{\prime}$ stands as its coupling constant. There is one corresponding real parameter in the algebra, asociated to the real gauge field $B^{0}_{\mu}$. Therefore, the $U(1)$ unitary irreps can be put as
\begin{equation}
U(\xi^{0})=\exp{(-ig^{\prime}YB^{0}_{\mu}(x))}
\end{equation}

For $SU(5)_{L}$ things are a little more complicated due to the fact that the Hermitian generators $T^{\dagger}_{a}=T_{a}$ of its Lie algebra $su(5)$ do not commute with one another. That is, they are subject to the canonical commutation relations of the form
\begin{equation}
\left[T_{a},T_{b}\right]=iC_{ab}^{c}T_{c}
\end{equation}
with $C_{ab}^{c}$ standing for the real and completely antisymmetric structure constants. For each $a,b,c,...$ there is a corresponding Hermitian generator out of the $n^{2}-1=24$ such Hermitian generators. Out of the $24$, only $4$ are diagonal (Cartan sub-algebra) and $20$ are off-diagonal.

The trace condition (that holds for every representation $\rho$) implies the normalization of the generators in the manner
\begin{equation}
\text{Tr}(T_{a})=0 \implies \text{Tr}(T_{a}T_{a})=\frac{1}{2} \delta_{ab} 
\end{equation}
  
The $SU(5)$ unitary irreducible representations can be put as
\begin{equation}
U(\xi)=\exp{(-igT_{a}A^{a}_{\mu}(x))}
\end{equation} 
where $g$ is the coupling constant assumed, throughout our proceedings here, to be identical with the SM coupling of $SU(2)_{L}$

Finally, we display the concrete expressions of the four diagonal Hermitian generators that are manipulated in our computations. They are in order: $D_{1}=T_{3}$, $D_{2}=T_{8}$, $D_{3}=T_{15}$, $D_{4}=T_{24}$.

\begin{equation}
T_{3}=\frac{1}{2}\left(\begin{array}{ccccc}
1 &  0 & 0 & 0 & 0 \\
0 & -1 & 0 & 0 & 0 \\
0 &  0 & 0 & 0 & 0 \\
0 &  0 & 0 & 0 & 0 \\
0 &  0 & 0 & 0 & 0 
\end{array}\right) \quad,\quad T_{8}=\frac{1}{2\sqrt{3}}\left(\begin{array}{ccccc}
1 & 0 &  0 & 0 & 0 \\
0 & 1 &  0 & 0 & 0 \\
0 & 0 & -2 & 0 & 0 \\
0 & 0 &  0 & 0 & 0 \\
0 & 0 &  0 & 0 & 0 
\end{array}\right) 
\end{equation}

\begin{equation}
T_{15}=\frac{1}{2\sqrt{6}}\left(\begin{array}{ccccc}
1 & 0 &  0 & 0 & 0 \\
0 & 1 &  0 & 0 & 0 \\
0 & 0 & 1 & 0 & 0 \\
0 & 0 &  0 & -1 & 0 \\
0 & 0 &  0 & 0 & 0 \end{array}\right) \quad,\quad T_{24}=\frac{1}{2\sqrt{10}}\left(\begin{array}{ccccc}
1 & 0 &  0 & 0 & 0 \\
0 & 1 &  0 & 0 & 0 \\
0 & 0 & 1 & 0 & 0 \\
0 & 0 &  0 & 1 & 0 \\
0 & 0 &  0 & 0 & -4 
\end{array}\right)
\end{equation}

The $SU(5)$ Hermitian generators of the group transform according to the adjoint representation of the $su(5)$ algebra
\begin{equation}
U(\xi)T_{a}U^{\dagger}(\xi)=\text{Adj}(\xi)^{b}_{a}T_{b}
\end{equation} 
where
\begin{equation}
\text{Adj}(\xi)=\exp(-ig\text{adj}(\xi))
\end{equation}
and
\begin{equation}
\text{adj}(\xi)=\xi^{a}\text{adj}(T_{a})
\end{equation}
which is exactly the adjoint representation of the algebra
\begin{equation}
\text{adj}(T_{a})^{b}_{c}=iC_{ac}^{b}
\end{equation}
We conclude our algebraical detour here with the manner the fermion representation transform under the gauge group of interest here. Denoting by $\psi^{\rho}$ any fermion multiplet (belonging to the unitary irreducible representation $\rho$ of the $SU(5)_{L}\times U(1)_{Y}$ gauge group), we have the transformation rule:
\begin{equation}
\psi^{\rho}(x)\rightarrow \exp [-i(gA^{\rho}_{\mu}(x)+g^{\prime}Y^{\rho}B_{\mu}(x))]\psi^{\rho}(x)
\end{equation} 
in order to obey the gauge symmetry of the model.
The gauge fields transform under the same gauge group in the manner:
\begin{equation}
A_{\mu}(x) \rightarrow U(\xi)A_{\mu}(x)U^{\dagger}(\xi)-\frac{i}{g}(\partial_{\mu}U(\xi))U^{\dagger}(\xi)\quad,\quad B_{\mu}(x) \rightarrow B_{\mu}(x)+\partial_{\mu}\xi^{0}(x)
\end{equation}

\section{Diagonalization of the boson mass matrix}

Here we enter into some technical details regarding the eigenvalues of the mass matrix \eqref{eq:6.13}. We start with the characteristic equation that reads  

\begin{equation}
\left|\begin{array}{cc}
\frac{1}{3}\left(1+\frac{7}{2}a-\frac{9}{2}b\right)-\frac{2a}{1-sin^{2}\theta_{W}} & \frac{1}{\sqrt{15}\cos\theta}\left(1-\frac{5}{2}a+\frac{3}{2}b\right)\\
\\
\frac{1}{\sqrt{15}\cos\theta}\left(1-\frac{5}{2}a+\frac{3}{2}b\right) & \frac{1}{5\cos^{2}\theta}\left(1+\frac{15}{2}a+\frac{15}{2}b\right)-\frac{2a}{1-sin^{2}\theta_{W}}
\end{array}\right|=0
\end{equation}
once it is enforced the well-known SM-condition $M^{2}(Z)=M^{2}(W^{\pm})/cos^{2}\theta_W$. This restriction will lead to a particular relation between the two free parameters - $a$ and $b$ - as the determinant is developed. It reduces to simply summing the following coeffcients of the powers of the parameters  

\begin{itemize}
\item free term: $\frac{1}{3} \left(\frac{1}{5\cos^{2}\theta}\right)-\left( \frac{1}{\sqrt{15}\cos\theta} \right)\left( \frac{1}{\sqrt{15}\cos\theta} \right)=0$
\item $a$: $-\frac{(5+7\sin^{2}\theta_W)}{2(1-\sin^{2}\theta_W)}-\frac{(5-17\sin^{2}\theta_W)}{2(1-\sin^{2}\theta_W)}+\frac{10(1-\sin^{2}\theta_W)}{2(1-\sin^{2}\theta_W)}=0$ 
\item $b$: $-\frac{9}{2}+\frac{15}{2}-2\frac{3}{2}=0 $
\item $ab$: $\frac{15(5+7\sin^{2}\theta_W)}{4(1-\sin^{2}\theta_W)}+\frac{9(5-17\sin^{2}\theta_W)}{4(1-\sin^{2}\theta_W)}+\frac{30(1-\sin^{2}\theta_W)}{4(1-\sin^{2}\theta_W)}=-\frac{-72\sin^{2}\theta_W}{1-\sin^{2}\theta_W}$
\item $a^2$: $\frac{(5+7\sin^{2}\theta_W)(5-17\sin^{2}\theta_W)}{4(1-\sin^{2}\theta_W)^{2}}-\frac{25}{4}=-\frac{-36\sin^{4}\theta_W}{(1-\sin^{2}\theta_W)^2}$
\item $b^2$: $-15\left(\frac{9}{4}\right)-\frac{9}{4}=-36$
\end{itemize}

Now, the characteristic equation becomes 

\begin{equation}
\left(a\tan^{2}\theta_W+b\right)^2=0
\end{equation}
and it ensures, up to a factor $\frac{m^{2}}{2}$, the following one-parameter mass matrix

\begin{equation}
\left(\begin{array}{cc}
\frac{1}{3}\left[1+a\frac{(7+2\sin^{2}\theta_{W})}{2(1-\sin^{2}\theta_{W})}\right] & \frac{1}{\sqrt{3(5-8\sin^{2}\theta_{W})}} \left[1-a\frac{(5-2\sin^{2}\theta_{W})}{2(1-\sin^{2}\theta_{W})}\right]\\
\\
\frac{1}{\sqrt{3(5-8\sin^{2}\theta_{W})}} \left[1-a\frac{(5-2\sin^{2}\theta_{W})}{2(1-\sin^{2}\theta_{W})}\right] & \frac{1}{5-8\sin^{2}\theta_{W}}\left[1+a\frac{(15-30\sin^{2}\theta_{W})}{2(1-\sin^{2}\theta_{W})}\right]
\end{array}\right)
\end{equation}
once $b$ was replaced by $-a\tan^{2}\theta_W$ in eq.\eqref{eq:6.13}. The trace of this matrix  will provide us with its eigenvalues $M_{1}=M^{2}(Z)$ and $M_{2}=M^{2}(Z^{\prime})$ where $M^{2}(Z^{\prime})=Tr(M^{2})-\frac{m^{2}a}{1-\sin^{2}\theta_W}$ and hence the concrete expression in eq.\eqref{eq:6.16}. 

It remains only to be determined the appropriate rotation matrix ($\omega$) that does the job of diagonalization, since it is a key ingredient in determining fermions' couplings to neutral bosons. However, this lies beyond the scope of our article here.

\end{document}